# Approaching the Practical Conductivity Limits of Aerosol Jet Printed Silver

Eva S. Rosker,* Michael T. Barako, Evan Nguyen, Don DiMarzio, Kim Kisslinger, Dah-Weih Duan, Rajinder Sandhu, Mark S. Goorsky, and Jesse Tice

**ABSTRACT:** Previous efforts to directly write conductive metals have been narrowly focused on nanoparticle ink suspensions that require aggressive sintering (>200 °C) and result in low-density, small-grained agglomerates with electrical conductivities <25% of bulk metal. Here, we demonstrate aerosol jet printing of a reactive ink solution and characterize high-density (93%) printed silver traces having near-bulk conductivity and grain sizes greater than the electron mean free path, while only requiring a low-temperature (80 °C) treatment. We have developed a predictive electronic transport model which correlates the microstructure to the measured conductivity and identifies a strategy to approach the practical conductivity limit for printed metals. Our analysis of how grain boundaries and tortuosity contribute to electrical resistivity provides insight into the basic materials science that governs how an ink formulator or process developer might approach improving the conductivity. Transmission line measurements validate that electrical properties are preserved up to 20 GHz, which demonstrates the utility of this technique for printed RF components. This work reveals a new method of producing robust printed electronics that retain the advantages of rapid prototyping and three-dimensional fabrication while achieving the performance necessary for success within the aerospace and communications industries.

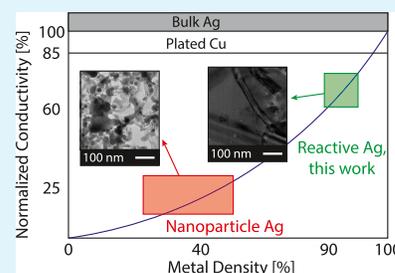

**KEYWORDS:** *aerosol jet printing, reactive metal ink, nanoparticle ink, printed electronics, electrical conductivity*

## ■ INTRODUCTION

Additively manufactured electronic components have experienced a rapid growth over the past decade[1−3] for prototyping and three-dimensional material deposition. However, printed metals often fail to achieve electrical conductivities comparable to conventionally microfabricated metals and are typically relegated to rapid prototyping applications rather than high-performance devices. Most commercially available silver inks are composed of silver nanoparticles, flakes, and/or other solids suspended in organic or aqueous solvents.[4−8] These suspension-based inks require physical transport of the metal-loaded ink to the substrate surface and require aggressive sintering and/or postprocessing to drive off the carrier liquids and bind the packed bed of particles into a continuous agglomerate. The sintering process is strongly temperature-dependent[9] and leads to undesired effects including trapped solvent, low density, tortuosity, and a large number of high-resistance interfaces that collectively result in low electrical conductivity (<25% of bulk metal). Printed nanoparticle traces are highly porous[10−12] and are significantly more resistive than conventionally manufactured metals such as sputtering, evaporation, and electroplating.[13] This is particularly significant for additively manufactured radio frequency (RF) applications that are sensitive to skin depth and require dense metals, low insertion losses, and an absence of parasitic interfaces that interfere with RF transmission. Postprocessing methods such as high-temperature baking (>200 °C), photonic curing, or pulsed laser sintering can incrementally increase conductivity[14−16] to within half that of bulk silver ($\sigma_{Ag}$ = 6.3 × 10$^7$ S/m), but these methods exceed the degradation temperature of most additively manufactured or polymeric substrates.

Throughout the past decade, a promising alternative to nanoparticle-based inks has emerged in the form of particle-free reactive metal inks, in which a metal organic decomposition (MOD) reaction converts a liquid precursor material into pure metal. The most ubiquitous approach, as originally described by Walker and Lewis, employs a modified Tollens' process in which silver acetate is dissolved in aqueous ammonium hydroxide and the resulting diamminesilver(I) complex is reduced to produce dense metallic silver.[17−19] A variety of other MOD reactions that utilize different silver-organo-complexes precursors have also been documented to yield solid silver with high conductivity.[20−24] Reactive inks avoid the inherent chemistry complications associated with

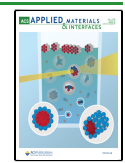

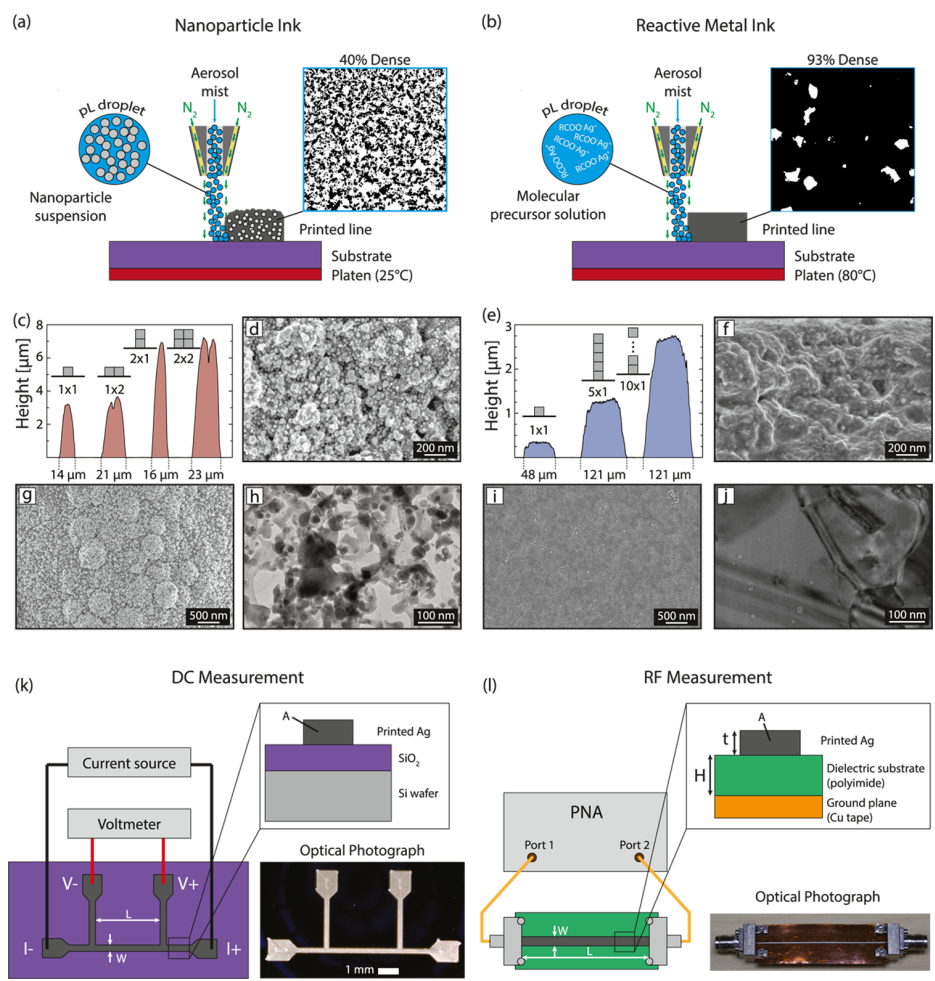

**Figure 1.** Representation of the printing process and resulting microstructures for the AJP nanoparticle (a) and reactive (b) inks, where voids are indicated by the white color contrast. Profilometry scans (c,e), cross-section SEM (d,f), plan-view SEM (g,i), and bright-field TEM (h,j) of nanoparticle and reactive metal traces. Experimental setup for DC (k) and RF (l) device tests.

surfactants and high-boiling-point organics present in nanoparticle suspensions and remove the need for high-temperature sintering. Like nanoparticle inks, reactive inks are compatible with direct-write technologies—the most common is drop-on-demand (DoD), but successful patterning of reactive inks has also been demonstrated with other methods including electrohydrodynamic printing[25] and spin-coating.[26] In addition to silver, the MOD approach has also been applied to other metal systems such as gold, copper, aluminum, and nickel.[27−32]

Highly conductive metals that can be patterned with commercially available tools and inks are prime candidates for implementation within the semiconductor industry. Circuit designers typically derate metals to ∼75% of the conductivity of a pure single-crystal metal because of the presence of defects and grain boundaries in plated metals, so we define this value to be the practical conductivity limit within the engineering context of this work. Here, we examine the conductivity and microstructure of silver nanoparticle inks and reactive silver inks directly written into metal traces using aerosol jet printing (AJP). AJP has been developed as a high-fidelity tool capable of printing a wide variety of materials at resolutions on the order of 10 μm line width,[33,34] but to date, it has been underrepresented as a tool for patterning reactive inks. With AJP, we transport the molecular precursor in a collimated spray onto a substrate surface held at 80 °C. The resulting wet solution is converted via rapid (∼1 s) thermal decomposition into a dense solid metal.[35] Figure 1 depicts a schematic representation of the comparison between nanoparticle (a) and reactive metal inks (b). Both inks are aerosolized into pL-sized droplets. Nanoparticle ink droplets contain solid metal suspended in the carrier solvent, while the reactive metal ink droplets contain only the solution-phase molecular precursor. After printing, the nanoparticle trace is observed to be a low-density (40%) network of agglomerated spheres where the nanoparticles are pulled into loose contact by surface energy forces and reinforced with sintering. In contrast, the reactive metal trace is observed to be a high-density (93%) series of overlapping large-grained metal plates that form a polycrystalline trace during thermal decomposition. The line profiles shown for the nanoparticle and reactive ink in Figure 1c,e demonstrate high-aspect-ratio traces with the ability to print multiple layers to add thickness to the trace. The scanning electron microscopy (SEM) and transmission electron microscopy (TEM) images in Figure 1d,f−j show the stark difference in microstructure between the traces. The nanoparticle trace is composed of many small particles (∼20 nm) that are loosely packed and only contact at the edges, whereas the reactive trace shows large plates (∼120 nm) that are densely packed with minimal voiding or discontinuities. The surface images also suggest that the reactive trace is smoother and more uniform than the nanoparticle trace.

**Device Design and Fabrication.** We print our devices with an Optomec AJ300 aerosol jet printer fitted with a 100 $\mu$m nozzle. The ink may be aerosolized either ultrasonically or pneumatically, and the printer is operated in a regime in which dry air acts as an annular sheath gas to contain the ink mist within a collimated beam. The ink is printed onto a substrate on a heated platen. Multiple layers are printed to achieve a range of aspect ratios to ensure that our measured conductivity is insensitive to device geometry and is derived primarily from intrinsic material properties. The DC conductivity devices are printed on a 500 $\mu$m-thick silicon wafer with 285 nm thermal $SiO_2$ that acts as a dielectric barrier to prevent current leakage. We print four-point device structures as shown in Figure 1k to eliminate the effect of contact resistance. We use high-aspect-ratio lines to increase the device resistance and improve our overall measurement sensitivity and signal-to-noise ratio. Stylus profilometry is used to determine the cross-sectional area of each trace in order to calculate the media conductivity from the measured resistance. To probe high-frequency RF properties (0.1−20 GHz), we print microstrip transmission lines as shown in Figure 1l onto a 140 $\mu$m-thick polyimide film and measure the transmission line S-parameters.

The silver nanoparticle traces are printed with a commercially available silver nanoparticle ink (Advanced Nano Products AS 1:4) using ultrasonic atomization. We use a nozzle-to-substrate distance of 3 mm, a sheath flow rate of 10 sccm, and an atomizer flow rate of 15 sccm to ensure we are printing continuous traces with minimal overspray. With a print head speed of 2 mm s$^{-1}$, this results in a single-trace width of 13.6 ± 0.2 $\mu$m. We print four aspect ratios: a single trace (1 × 1), two traces printed side-by-side (1 × 2), a single trace printed with two passes (2 × 1), and a double trace with two passes (2 × 2). After printing, the devices are sintered for 2 h in a 200 °C oven as specified by the ink manufacturer. The reactive ink used in this study (Electroninks EI-1403) was synthesized using the modified Tollens' chemistry described previously. The traces are printed using pneumatic atomization, in which we use a nozzle-to-substrate distance of 3 mm, a sheath flow rate of 70 sccm, an atomizer flow rate of 500 sccm, and an exhaust flow rate of 540 sccm. With a print head speed of 5 mm s$^{-1}$, this results in a continuous wet film that is 50 ± 5 $\mu$m wide and devoid of any overspray effects. The platen is heated to maintain a substrate temperature of 80 °C to enable thermal decomposition of the liquid ink upon printing. Minimal spreading is observed in the moment between deposition and reaction, which affects the achievable line width. We vary the aspect ratio of the traces by maintaining a constant target width and increasing the number of printer passes to build up the material thickness. A single pass deposits a peak trace height of 0.35 ± 0.07 $\mu$m, and the final film thickness scales approximately linearly with the number of passes (see Figure S2 in the Supporting Information). The four-point structures are printed from 1 to 10 passes (0.35−3.5 $\mu$m, respectively) to ensure that the material properties are independent of thickness or near-interface effects. The RF microstrips are printed 10 layers thick to ensure a metal thickness greater than the skin depth for RF in the frequency range of our tests. Although the reactive traces are fully solidified after printing on the heated platen, we also explore longer annealing times (0.5, 1, and 25 h) to assess the asymptotic limit to conductivity. A second set of four-point structures was annealed for 1 h at 300 °C to test the effects of very high-temperature annealing.

The method of ink deposition can also have an effect on the printed metal morphology. DoD printing generally produces smoother and higher density traces,[4] so we also characterize DoD reactive metal traces to compare the resulting microstructure to that of our AJP reactive traces. We use a Fujifilm Dimatix 2831 equipped with a 21.5 $\mu$m nozzle. To expel the ink droplets (∼10 pL volume), we use a standard Dimatix Model Fluid waveform with a voltage of 22 V, a duration of 3.5 $\mu$s, and a frequency of 2.5 kHz. During printing the platen was maintained at 60 °C, and after printing the traces underwent a 160−210 °C ramp annealing to drive off any residual organic solvent.

**Printed Metal Morphology.** The electrical properties of printed metals depend on a combination of the intrinsic material chemistry and the morphology (see Table S2 in the Supporting Information), which includes contributions from the porosity and voiding, the tortuosity that artificially increases the conduction path length, the interfacial topology that increases surface scattering (particularly at high frequencies), and the grain size and grain distribution which comprise the microstructure.

We characterize the microstructure and porosity of our printed metal traces using a combination of SEM, TEM, and X-ray diffraction (XRD). The material microstructure contributes to electron scattering from interfaces and other internal defects. XRD and TEM selected area diffraction reveal that the nanoparticle grains are randomly oriented, which is to be expected for an agglomeration of spherical particles. We calculate the average grain size to be 24 nm using the Scherrer equation,[36] which is in good agreement with the measured average line-of-sight distance of 21 nm observed from TEM image processing. The AJP reactive traces demonstrate slight (111) texturing which is likely because of preferential formation of close-packed faces during crystal growth immediately after printing.[37] TEM reveals a much larger average line-of-sight distance of 120 nm for the AJP traces and 460 nm for the DoD traces. Additionally, energy-dispersive spectroscopy shows the reactive ink traces to be nearly pure silver. This was corroborated by FTIR (Thermo Scientific Nicolet 6700 FTIR equipped with a Golden Gate ATR crystal) in which no organic contamination was detected above the noise floor of the measurement.

To obtain the density of our printed traces ($\Phi_{solid}$), we use image processing of TEM images and assume that the two-dimensional slice is representative of a randomly oriented, isotropic material. After thresholding and binarizing the images, we perform a particle analysis to yield $\Phi_{solid}$ = 40% ± 6% solid for the nanoparticle traces, $\Phi_{solid}$ = 93% ± 2% solid for the AJP reactive silver, and $\Phi_{solid}$ = 95% ± 1% for the DoD reactive silver. The relatively large (60%) void volume in the nanoparticle traces is attributed to the low solid content of the packed nanoparticles in the printed film. The small (7 and 5%) void volume in the reactive metal traces is likely formed during the evaporation of the organic solvent during printing. A material's density can affect its conductivity in multiple ways. First, the current density is increased by an increase in conduction path volume. A less dense material such as the nanoparticle traces contains a reduced area through which it can conduct electrons. Second, the tortuosity of the electrical pathway due to low interface area artificially lengthens the path an electron must travel. Finally, the large number of discrete particles increases surface and grain boundary scattering, the former of which is particularly affected at higher frequencies. Compared to nanoparticle inks, reactive inks possess significantly larger grains, reduced porosity, and a lack of chemical impurities. Combined, these factors are the reason we observe such a dramatic improvement in electrical properties which exceed other nanoparticle-based printing efforts to date.

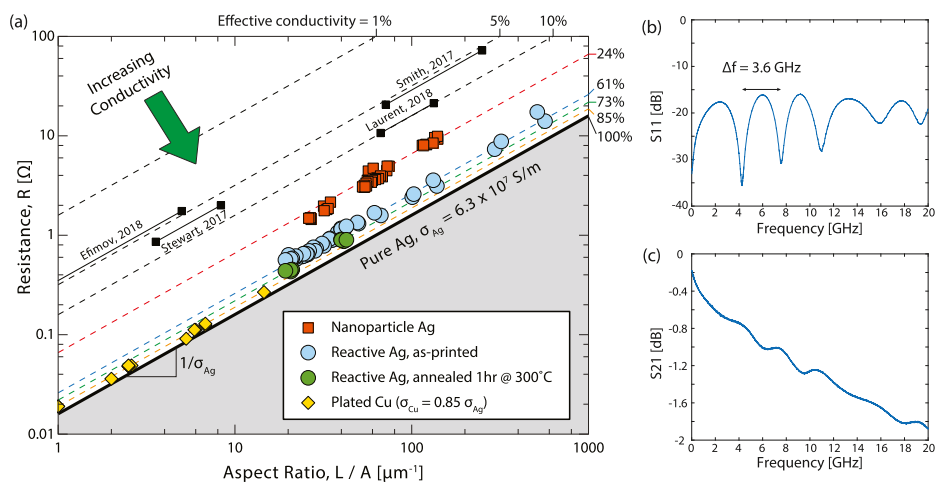

**Figure 2.** (a) Conductivity of AJP nanoparticle and reactive silver traces. Reactive inks are more conductive than nanoparticle inks and even approach the conductivity of plated copper, a well-known industry standard. Reference data reproduced from refs 42−45. (b,c) RF S-parameters measured for a set of the as-printed reactive ink microstrips show features characteristic of a highly conductive metal.

**DC Conductivity.** We measure the resistance of the printed traces by sweeping current from −10 to +10 mA through the device and measuring the voltage across the test section. The resistance is calculated from the slope of a linear regression fit to the I−V characteristic. We calculate the media conductivity ($\sigma$) by measuring the resistance of lines with different aspect ratios (equal to length $L$ divided by cross-sectional area $A$, see Figure 2a), and fitting the equation

$$R = \frac{1}{\sigma}\left[\frac{L}{A}\right] \quad (1)$$

Figure 2a shows the resistance of annealed, printed nanoparticle traces with a best fit conductivity value equal to 24% of pure bulk silver and the resistance of the as-printed (i.e., no annealing) reactive ink traces with a best fit conductivity value equal to 61% of bulk silver. The reactive ink trace conductivity remained unchanged after annealing in a 100 °C oven for 1 h, which is expected because the traces already reacted during printing and small quantities of additional heat do not provide enough thermal energy to appreciably alter the microstructure via grain growth or diffusion. However, the addition of a postprocess annealing step at high temperature may enhance conductivity via grain ripening.[38,39] After a 1 h bake in a 300 °C oven, the conductivity increased to 73% of bulk silver, which is denoted by the green data points. As a comparison, electroplated copper traces used widely in the industry[40,41] are observed to have a conductivity of 90% of bulk copper, which is equivalent to ∼85% of bulk silver.

For a polycrystalline porous metal, the maximum achievable conductivity is limited by the diffusive-transport effects associated with the metal's porosity. We use Maxwell's effective medium theory (EMT) to estimate the conductivity reduction for subunity densities in isotropic porous media.[46] This model is most accurate for high-density porous metals such as those found in our reactive ink traces and offers a meaningful approximation for the low-density porous metals found in our nanoparticle ink traces. Because the voids do not contribute to conduction, the effective conductivity of a medium due to reduced density $\sigma_{\text{eff,EMT}}$ can be written as

$$\sigma_{\text{eff,EMT}} = \sigma_{\text{metal}}\left(\frac{2\Phi_{\text{solid}}}{3 - \Phi_{\text{solid}}}\right) \quad (2)$$

where $\Phi_{\text{solid}}$ is the volume fraction of the conductive media (in which $\Phi_{\text{solid}} = 1 - \Phi_{\text{void}}$), and $\sigma_{\text{metal}}$ is the bulk conductivity of the metal phase (i.e., $6.3 \times 10^7$ S/m for silver). This expression is valid in the limit of a dilute and even distribution of spherical voids embedded in solid material.

In addition to transport limitations due to porosity, the subcontinuum effects of grain boundary scattering are present when the grain size is comparable to the intrinsic electron mean free path $\lambda_o$. We consider the length scales of the morphological features in our devices using Matthiessen's rule[47] to define the effective mean free path $\lambda_{\text{eff}}$

$$\frac{1}{\lambda_{\text{eff}}} = \frac{1}{\lambda_o} + \frac{1}{\lambda_{\text{LoS}}} \quad (3)$$

where $\lambda_o$ is the mean free path of an electron in silver (53 nm at 300 K),[48] and $\lambda_{\text{LoS}}$ is the characteristic line-of-sight distance for an electron moving across a grain in the ballistic transport regime. In geometric terms, $\lambda_{\text{LoS}}$ represents the average chord length in a sphere of a given radius $r$ and is equal to $4r/\pi$. With TEM, we observe a two-dimensional cross-section of randomly oriented grains and measure this value with appropriate image analysis (see Figure S4 in the Supporting Information for details). This measured characteristic length is not necessarily the maximum distance across the spherical grain but is instead a representation of the distribution of paths that would cross the grain in three-dimensional space. Table 1 below presents the characteristic length and the estimated average grain size determined from this relationship.

According to kinetic theory,[49,50] our model assumes that there is a linear relationship between the electrical conductivity of a material and the corresponding electron mean free path

**Table 1. Characteristic Length and Estimated Grain Size from TEM**

|  | characteristic length $\lambda_{LoS}$ [nm] (measured from TEM image analysis) | average grain diameter [nm] (estimated from $4r/\pi$ relationship) |
|---|---|---|
| AJP nanoparticle | 21 | 33 |
| AJP reactive | 120 | 190 |
| DoD reactive | 460 | 720 |

$$\frac{\sigma_{eff,sc}}{\sigma_{metal}} = \frac{\lambda_{eff}}{\lambda_o} \quad (4)$$

Combining the previous equations yields the following expression for our multiscale model in which the effective conductivity is a function of macroscale diffusive transport effects as well as microscale subcontinuum effects

$$\frac{\sigma_{eff}}{\sigma_{metal}} = \frac{\lambda_o \lambda_{gb}}{\lambda_o(\lambda_o + \lambda_{gb})} \times \left[\left(\frac{2\Phi_{solid}}{3 - \Phi_{solid}}\right)\right] \quad (5)$$

Table 2 displays the material porosity, maximum theoretical conductivity, model-predicted conductivity, and experimentally determined conductivity for each of the metal traces in this work. While the model prediction is sufficiently close to the observed behavior for the reactive inks, the large discrepancy seen for the nanoparticle traces may be attributed to the inaccuracy of Maxwell's EMT at low densities.[51] At such low volume fractions, the assumptions present in EMT (spherical, spatially separated voids) are not met and will result in an inaccurate prediction. It is also possible that our model overpredicts the impact of grain boundary scattering. Matthiessen's rule is a scaling argument that is best used for estimating the effects of independent scattering mechanisms in a complex medium, and it becomes sensitive to uncertainty when there are multiple dominant scattering mechanisms. As the complexity of the structure increases, multiple scattering rates become comparable in magnitude, and this approximation may not provide a precise prediction of total conductivity. This could come from a subunity scattering rate at the grain boundaries.[52] Because the reactive inks have grains much larger than the mean free path, Matthiessen's rule can be thought of as a perturbation from bulk transport. However, the small grains of nanoparticle inks lead to a scenario where the competing scattering mechanisms are all comparable and any uncertainty in one input (e.g. grain size) may lead to a large error in the predicted quantity, as seen here.

The predicted effective conductivity versus characteristic length determined from the two-dimensional TEM image is shown in Figure 3a, and the statistical distribution shown in Figure 3b demonstrates the significant difference in grain size for the nanoparticle versus AJP and DoD reactive metal traces. This distribution was created using a number-weighted average; the volume-weighted average is displayed in the inset to accurately reflect the three-dimensional transport occurring in the materials system. The AJP reactive silver possesses an average characteristic length nearly an order of magnitude larger than that of the nanoparticle traces, and it is even larger for the DoD reactive silver traces. The larger grains in the reactive ink traces and longer line-of-sight conduction paths between voids compared the nanoparticle traces increase the effective mean free path toward the bulk value with a corresponding preservation of bulk-like conductivity. The difference in grain size between AJP and DoD traces may be attributed to the underlying nucleation and growth mechanisms within each deposition method.

This model describes the path to approach the practical conductivity limit of ~75% through thoughtful engineering of the materials system. To advance beyond this limit requires adjustments in pure metallurgy and single-crystal growth, both of which are significantly more niche aspects of material fabrication than conventional device processing or electronics manufacturing. One method will be to increase the density of the printed metal. As trace density approaches unity, the EMT limit approaches that of a bulk metal. The conductivity can also be improved by increasing the grain size of the printed metal. Larger grains result in fewer grain boundaries to act as scattering sites, which is expressed in the trend toward the EMT limit at length scales well above the electron mean free path. Because the grain size of nanoparticle traces is ultimately limited by the particle size in the original ink suspension, reactive inks present a unique opportunity to reach an area of processing space that has previously been inaccessible. Notably, this is made possible through the use of commercially available tools and inks as opposed to the specialized equipment or modifications that would be difficult to reproduce in a semiconductor foundry setting.

**RF Measurements.** At radio frequencies, the volume of available conductor is dictated by the skin depth effect,[53] in which electrons migrate toward the surface and do not travel uniformly within the bulk volume of the trace as in the DC limit. The skin depth is the distance from the conductor edge where the signal diminishes to $1/e$ of the surface conductivity, which is highly dependent on frequency. Therefore, a smooth, dense metal that is free from significant porosity is required at high frequencies to efficiently conduct current. We measure the material properties of the reactive metal traces from 0.1 to 20 GHz by printing 50 Ω microstrip transmission lines and using a network analyzer to extract the S-parameter matrix. This quantifies the relative amount of energy that is transmitted through the signal line and the associated losses from reflections at the interfaces between the line and the connectors. A connectorized microstrip test setup is compatible with a variety of standardized RF testing equipment and coaxial calibration kits, which allows us to rapidly assess initial performance of the printed device properties. The microstrip device design shown in Figure 1l is printed on a 140 μm-thick polyimide substrate with copper tape affixed to the back side as the ground metal. The devices are connectorized with two 2.4 mm jack end launch connectors (Southwest Microwave model

**Table 2. Predicted and Measured Conductivity Values**

|  | $\Phi_{metal}$ ($\Phi_{void}$) metal density (void fraction) | $\sigma_{eff,EMT}$ EMT conductivity limit (from eq 2) (%) | $\sigma_{eff}$ predicted conductivity (from eq 5) (%) | $\sigma_{meas}$ experimentally measured conductivity (%) |
|---|---|---|---|---|
| AJP nanoparticle | 0.40 (0.60) | 31 | 10 | 24 |
| AJP reactive | 0.93 (0.7) | 90 | 63 | 61 |
| DoD reactive | 0.95 (0.5) | 93 | 83 | 82 |

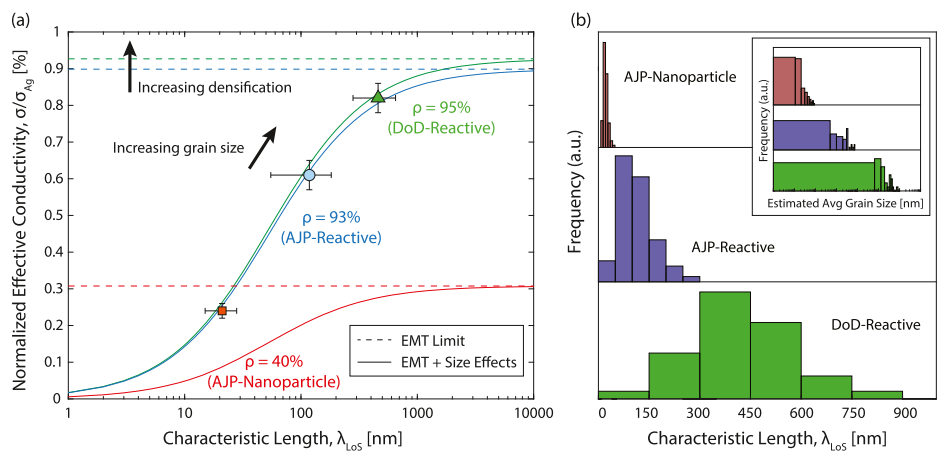

**Figure 3.** (a) Predictive model for determining the conductivity of the printed metal with experimental data shown. There is an inherent limit based on the porosity of the material, which is further refined by the average grain size. (b) The distribution of characteristic lengths for each material used in this work. The main figure displays the distribution as a number-weighted average, while the inset shows a volume-weighted average of the estimated grain size calculated from the characteristic length. The mean grain size for both of these methods is the same value.

no. 1492-02A-6) and calibrated cables driven by a programmable network analyzer (PNA).

The signal reflection and transmission of the as-printed reactive ink microstrips are shown in Figure 2b,c, respectively. The lobe pattern present in the measured return loss ($S_{11}$) is consistent with a simulation created from known physical properties of the device including the transmission line width (310 $\mu$m), length (2.5 cm), metal thickness (3 $\mu$m), and substrate dielectric constant (3.5). We confirm our measurement by using the following equation to calculate the length of the transmission line from the periodicity of the lobes

$$L = \frac{c}{2\sqrt{\varepsilon_r} \times \Delta f} \tag{6}$$

where $L$ is the length of the line, $c$ is the speed of light in vacuum, $\varepsilon_r$ is the dielectric constant of the polyimide substrate, and $\Delta f$ is the spacing between the periodic lobes ($\Delta f$ = 3.6 GHz). The calculated length is 2.3 cm, which matches the measured length to within ∼8%. The agreement between the measured and simulated data and corroboration across measurements demonstrates that the reactive ink trace has electrical transport properties similar to a conventional metal up to radio frequencies. Furthermore, the return loss for the reactive traces is low (<15 dB) across the entire range of frequencies tested, which is characteristic of a dense conductive metal that has a smooth interface with the substrate. The magnitude of the return loss includes contributions from reflection at the connector interface that cannot be independently isolated using this measurement setup and necessitates more specialized RF testing techniques. The $S_{21}$ measurements are consistent with low-frequency simulations and confirm the material conductivity values obtained from DC four-point testing. The shape of the curve is also consistent with that of a transmission line, which has a steeper slope proportional to $1/f$ at low frequencies, but transitions to a flatter slope with $1/\sqrt{f}$ scaling at higher frequencies (>2 GHz) because of the skin depth effect.

The performance of dense reactive metals is superior to that of porous nanoparticle traces in the RF regime as well as the DC limit. Printed nanoparticle lines performed so poorly in RF tests that the output data is functionally equivalent to an open circuit with signals that are impossible to fit to any physical model and are not reproducible, even among repeated tests of the same device. This was determined from auxiliary experiments to originate from a combination of poor contact at the connector interface as well as the intrinsic discontinuities and reflections within the low-density metal, and the deconvolution of these two effects is otherwise impossible.

Dense, highly conductive metals directly translate to enhanced performance in RF devices. This RF demonstration shows that reactive metal traces can be printed to be virtually indistinguishable from conventionally microfabricated metals from the DC limit to 20 GHz. Printable reactive metal inks could feasibly replace conventional metals for a variety of RF devices including waveguides and antennas and can enable three-dimensional RF structures that cannot be conventionally manufactured. This approach is well-suited for a variety of high-performance device applications and related RF components within the aerospace industry.

## ■ CONCLUSIONS

We used AJP to fabricate conductive metal traces from nanoparticle and reactive inks to contrast their microstructures and conductivities and to provide an initial demonstration of RF performance. The DC conductivity values are supported by a simple kinetic theory electrical transport model that accurately predicts the observed conductivity to within 3% error for the reactive metal traces. We identify a path through densification and grain refinement to approach the practical conductivity limits (∼75%) using printed reactive metal inks. As we approach the ideal morphology of printed metal traces, further process optimization can lead to larger grain sizes and fewer voids to achieve higher electrical conductivity. Density can be increased by reducing the solvent evaporation rate through temperature control during solidification of the wet film and through refinement in ink chemistry. Alternative printing methods which reduce the number of nucleation sites will also lead to larger grain size and thus higher conductivity. This work represents a critical path toward high-performance additively manufactured electronics and RF components.

## ■ ASSOCIATED CONTENT

**Supporting Information**

The Supporting Information is available free of charge at https://pubs.acs.org/doi/10.1021/acsami.0c06959.

Toolpath files used to print the devices; printer output and printed metal features for the AJP reactive ink traces; TEM image and EDS analysis of an AJP reactive ink trace; TEM image processing technique; XRD results for the AJP nanoparticle and reactive ink traces; device design parameters; and factors affecting the conductivity of printed metals (PDF)

## ■ AUTHOR INFORMATION

**Corresponding Author**

Eva S. Rosker − NG Next, Northrop Grumman Corporation, Redondo Beach, California 90278, United States; UCLA Department of Materials Science & Engineering, Los Angeles, California 90095, United States; orcid.org/0000-0003-0990-4412; Email: erosker@ucla.edu

**Authors**

Michael T. Barako − NG Next, Northrop Grumman Corporation, Redondo Beach, California 90278, United States

Evan Nguyen − NG Next, Northrop Grumman Corporation, Redondo Beach, California 90278, United States

Don DiMarzio − NG Next, Northrop Grumman Corporation, Redondo Beach, California 90278, United States; Center for Functional Nanomaterials, Brookhaven National Laboratory, Upton, New York 11973, United States

Kim Kisslinger − Center for Functional Nanomaterials, Brookhaven National Laboratory, Upton, New York 11973, United States

Dah-Weih Duan − NG Next, Northrop Grumman Corporation, Redondo Beach, California 90278, United States

Rajinder Sandhu − NG Next, Northrop Grumman Corporation, Redondo Beach, California 90278, United States

Mark S. Goorsky − UCLA Department of Materials Science & Engineering, Los Angeles, California 90095, United States

Jesse Tice − NG Next, Northrop Grumman Corporation, Redondo Beach, California 90278, United States

Complete contact information is available at:
https://pubs.acs.org/10.1021/acsami.0c06959

**Notes**

The authors declare no competing financial interest.

## ■ ACKNOWLEDGMENTS

The electron microscopy work at the Center for Functional Nanomaterials, BNL was supported by the U.S. Department of Energy, Office of Basic Energy Science (BES), Scientific User Facilities Division, under Contract no. DE-SC0012704. The authors thank Jackson Ng, Stanley Fletcher, Stephane Larouche, and Vesna Radisic for their help with RF device design and testing. The authors also thank Electroninks Inc. for supplying the reactive silver ink used in this work.

## ■ REFERENCES

(1) Rosker, E. S.; Sandhu, R.; Hester, J.; Goorsky, M. S.; Tice, J. Printable Materials for the Realization of High Performance RF Components: Challenges and Opportunities. *Int. J. Antennas Propag.* **2018**, *2018*, 9359528.

(2) Huang, Q.; Zhu, Y. Printing Conductive Nanomaterials for Flexible and Stretchable Electronics: A Review of Materials, Processes, and Applications. *Adv. Mater. Technol.* **2019**, *4*, 1800546.

(3) Mckerricher, G.; Vaseem, M.; Shamim, A. Fully Inkjet-Printed Microwave Passive Electronics. *Microsystems Nanoeng.* **2017**, *3*, 16075.

(4) Seifert, T.; Sowade, E.; Roscher, F.; Wiemer, M.; Gessner, T.; Baumann, R. R. Additive Manufacturing Technologies Compared: Morphology of Deposits of Silver Ink Using Inkjet and Aerosol Jet Printing. *Ind. Eng. Chem. Res.* **2015**, *54*, 769−779.

(5) Agarwala, S.; Goh, G. L.; Yeong, W. Y. Optimizing Aerosol Jet Printing Process of Silver Ink for Printed Electronics. *IOP Conf. Ser. Mater. Sci. Eng.* **2017**, *191*, 012027.

(6) Mahajan, A.; Frisbie, C. D.; Francis, L. F. Optimization of Aerosol Jet Printing for High-Resolution, High-Aspect Ratio Silver Lines. *ACS Appl. Mater. Interfaces* **2013**, *5*, 4856−4864.

(7) Rajan, K.; Roppolo, I.; Chiappone, A.; Bocchini, S.; Perrone, D.; Chiolerio, A. Silver Nanoparticle Ink Technology: State of the Art. *Nanotechnol. Sci. Appl.* **2016**, *9*, 1−13.

(8) Nayak, L.; Mohanty, S.; Nayak, S. K.; Ramadoss, A. A Review on Inkjet Printing of Nanoparticle Inks for Flexible Electronics. *J. Mater. Chem. C* **2019**, *7*, 8771−8795.

(9) Volkman, S. K.; Yin, S.; Bakhishev, T.; Puntambekar, K.; Subramanian, V.; Toney, M. F. Mechanistic Studies on Sintering of Silver Nanoparticles. *Chem. Mater.* **2011**, *23*, 4634−4640.

(10) Xu, B. *Inkjet Printing of Silver for Direct Write Applications*; University of Manchester, 2010.

(11) Shen, W.; Zhang, X.; Huang, Q.; Xu, Q.; Song, W. Preparation of solid silver nanoparticles for inkjet printed flexible electronics with high conductivity. *Nanoscale* **2014**, *6*, 1622−1628.

(12) Vaithilingam, J.; Saleh, E.; Körner, L.; Wildman, R. D.; Hague, R. J. M.; Leach, R. K.; Tuck, C. J. 3-Dimensional Inkjet Printing of Macro Structures from Silver Nanoparticles. *Mater. Des.* **2018**, *139*, 81−88.

(13) Hara, T.; Shimura, Y.; Toida, H. Deposition of Low Resistivity Copper Conductive Layers by Electroplating from a Copper Hexafluorosilicate Solution. *Electrochem. Solid-State Lett.* **2003**, *6*, 97−99.

(14) Hwang, H.-J.; Oh, K.-H.; Kim, H.-S. All-photonic drying and sintering process via flash white light combined with deep-UV and near-infrared irradiation for highly conductive copper nano-ink. *Sci. Rep.* **2016**, *6*, 19696.

(15) Kang, J. S.; Ryu, J.; Kim, H. S.; Hahn, H. T. Sintering of Inkjet-Printed Silver Nanoparticles at Room Temperature Using Intense Pulsed Light. *J. Electron. Mater.* **2011**, *40*, 2268−2277.

(16) Hwang, J.-Y.; Moon, S.-J. The Characteristic Variations of Inkjet-Printed Silver Nanoparticle Ink During Furnace Sintering. *J. Nanosci. Nanotechnol.* **2013**, *13*, 6145−6149.

(17) Walker, S. B.; Lewis, J. A. Reactive Silver Inks for Patterning High-Conductivity Features at Mild Temperatures. *J. Am. Chem. Soc.* **2012**, *134*, 1419−1421.

(18) Zhao, Z.; Mamidanna, A.; Lefky, C.; Hildreth, O.; Alford, T. L. A Percolative Approach to Investigate Electromigration Failure in Printed Ag Structures. *J. Appl. Phys.* **2016**, *120*, 125104.

(19) Kastner, J.; Faury, T.; Außerhuber, H. M.; Obermüller, T.; Leichtfried, H.; Haslinger, M. J.; Liftinger, E.; Innerlohinger, J.; Gnatiuk, I.; Holzinger, D.; Lederer, T. Silver-Based Reactive Ink for Inkjet-Printing of Conductive Lines on Textiles. *Microelectron. Eng.* **2017**, *176*, 84−88.

(20) Jahn, S. F.; Blaudeck, T.; Baumann, R. R.; Jakob, A.; Ecorchard, P.; Rüffer, T.; Lang, H.; Schmidt, P. Inkjet Printing of Conductive Silver Patterns by Using the First Aqueous Particle-Free MOD Ink without Additional Stabilizing Ligands†. *Chem. Mater.* **2010**, *22*, 3067−3071.

(21) Wu, J.-T.; Hsu, S. L.-C.; Tsai, M.-H.; Hwang, W.-S. Inkjet Printing of Low-Temperature Cured Silver Patterns by Using AgNO3/1-Dimethylamino-2-propanol Inks on Polymer Substrates. *J. Phys. Chem. C* **2011**, *115*, 10940−10945.

(22) Chang, Y.; Wang, D.-Y.; Tai, Y.-L.; Yang, Z.-G. Preparation, Characterization and Reaction Mechanism of a Novel Silver-Organic Conductive Ink. *J. Mater. Chem.* **2012**, *22*, 25296−25301.

(23) Dong, Y.; Li, X.; Liu, S.; Zhu, Q.; Zhang, M.; Li, J.-G.; Sun, X. Optimizing Formulations of Silver Organic Decomposition Ink for Producing Highly-Conductive Features on Flexible Substrates: The Case Study of Amines. *Thin Solid Films* **2016**, *616*, 635−642.

(24) Black, K.; Singh, J.; Mehta, D.; Sung, S.; Sutcliffe, C. J.; Chalker, P. R. Silver Ink Formulations for Sinter-Free Printing of Conductive Films. *Sci. Rep.* **2016**, *6*, 1−7.

(25) Lefky, C.; Arnold, G.; Hildreth, O. High-Resolution Electrohydrodynamic Printing of Silver Reactive Inks. *MRS Adv.* **2016**, *1*, 2409−2414.

(26) Bhat, K. S.; Ahmad, R.; Wang, Y.; Hahn, Y.-B. Low-Temperature Sintering of Highly Conductive Silver Ink for Flexible Electronics. *J. Mater. Chem. C* **2016**, *4*, 8522−8527.

(27) Schoner, C.; Tuchscherer, A.; Blaudeck, T.; Jahn, S. F.; Baumann, R. R.; Lang, H. Particle-Free Gold Metal-Organic Decomposition Ink for Inkjet Printing of Gold Structures. *Thin Solid Films* **2013**, *531*, 147−151.

(28) Rosen, Y.; Grouchko, M.; Magdassi, S. Printing a Self-Reducing Copper Precursor on 2D and 3D Objects to Yield Copper Patterns with 50% Copper's Bulk Conductivity. *Adv. Mater. Interfaces* **2015**, *2*, 1400448.

(29) Farraj, Y.; Grouchko, M.; Magdassi, S. Self-Reduction of a Copper Complex MOD Ink for Inkjet Printing Conductive Patterns on Plastics. *Chem. Commun.* **2015**, *51*, 1587−1590.

(30) Shin, D.-H.; Woo, S.; Yem, H.; Cha, M.; Cho, S.; Kang, M.; Jeong, S.; Kim, Y.; Kang, K.; Piao, Y. A Self-Reducible and Alcohol-Soluble Copper-Based Metal-Organic Decomposition Ink for Printed Electronics. *ACS Appl. Mater. Interfaces* **2014**, *6*, 3312−3319.

(31) Murray, A. K.; Isik, T.; Ortalan, V.; Gunduz, I. E.; Son, S. F.; Chiu, G. T.-C.; Rhoads, J. F. Two-Component Additive Manufacturing of Nanothermite Structures via Reactive Inkjet Printing. *J. Appl. Phys.* **2017**, *122*, 184901.

(32) Li, D.; Sutton, D.; Burgess, A.; Graham, D.; Calvert, P. D. Conductive Copper and Nickel Lines via Reactive Inkjet Printing. *J. Mater. Chem.* **2009**, *19*, 3719−3724.

(33) Secor, E. B. Principles of Aerosol Jet Printing. *Flex. Print. Electron.* **2018**, *3*, 035002.

(34) Secor, E. B. Guided Ink and Process Design for Aerosol Jet Printing Based on Annular Drying Effects. *Flex. Print. Electron.* **2018**, *3*, 035007.

(35) Walker, S. *Synthesis and Patterning of Reactive Silver Inks*; University of Illinois at Urbana-Champaign, 2013.

(36) Scherrer, P. Bestimmung Der Größe Und Der Inneren Struktur von Kolloidteilchen Mittels Röntgenstrahlen. *Gottinger Nachrichten* **1918**, *2*, 98−100.

(37) Wang, S. G.; Tian, E. K.; Lung, C. W. Surface Energy of Arbitrary Crystal Plane of Bcc and Fcc Metals. *J. Phys. Chem. Solids* **2000**, *61*, 1295−1300.

(38) Park, J.-W.; Baek, S.-G. Thermal Behavior of Direct-Printed Lines of Silver Nanoparticles. *Scr. Mater.* **2006**, *55*, 1139−1142.

(39) Rahman, M. T.; McCloy, J.; Ramana, C. V.; Panat, R. Structure, Electrical Characteristics, and High-Temperature Stability of Aerosol Jet Printed Silver Nanoparticle Films. *J. Appl. Phys.* **2016**, *120*, 075305.

(40) Ahn, K. Y.; Forbes, L. Selective Electroless-Plated Copper Metallization. U.S. Patent 7,262,505B2, 2007.

(41) Lane, M. W.; Murray, C. E.; McFeely, F. R.; Vereecken, P. M.; Rosenberg, R. Liner Materials for Direct Electrodeposition of Cu. *Appl. Phys. Lett.* **2003**, *83*, 2330−2332.

(42) Efimov, A. A.; Arsenov, P. V.; Protas, N. V.; Minkov, K. N.; Urazov, M. N.; Ivanov, V. V. Dry aerosol jet printing of conductive silver lines on a heated silicon substrate. *IOP Conf. Ser. Mater. Sci. Eng.* **2018**, *307*, 012082.

(43) Stewart, I. E.; Kim, M. J.; Wiley, B. J. Effect of Morphology on the Electrical Resistivity of Silver Nanostructure Films. *Appl. Mater. Interfaces* **2017**, *9*, 1870−1876.

(44) Laurent, P.; Stoukatch, S.; Dupont, F.; Kraft, M. Electrical Characterization of Aerosol Jet Printing (AJP) Deposited Conductive Silver Tracks on Organic Materials. *Microelectron. Eng.* **2018**, *197*, 67−75.

(45) Smith, M.; Choi, Y. S.; Boughey, C.; Kar-narayan, S. Controlling and Assessing the Quality of Aerosol Jet Printed Features for Large Area and Flexible Electronics. *Flex. Print. Electron.* **2017**, *2*, 015004.

(46) Maxwell, J. *A Treatise on Electricity and Magnetism*, 3rd ed.; Oxford University Press: Oxford, 1892.

(47) Matthiessen, A.; Vogt, C. On the Influence of Temperature on the Electric Conducting Power of Alloys. *Phil. Trans. R. Soc.* **1864**, *154*, 167−200.

(48) Gall, D. Electron Mean Free Path in Elemental Metals. *J. Appl. Phys.* **2016**, *119*, 085101.

(49) Jain, A.; McGaughey, A. J. H. Thermal Transport by Phonons and Electrons in Aluminum, Silver, and Gold from First Principles. *Phys. Rev. B* **2016**, *93*, 081206.

(50) Barako, M. T.; Sood, A.; Zhang, C.; Wang, J.; Kodama, T.; Asheghi, M.; Zheng, X.; Braun, P. V.; Goodson, K. E. Quasi-Ballistic Electronic Thermal Conduction in Metal Inverse Opals. *Nano Lett.* **2016**, *16*, 2754−2761.

(51) Datta, S.; Chan, C. T.; Ho, K. M.; Soukoulis, C. M. Effective Dielectric Constant of Periodic Composite Structures. *Phys. Rev. B* **1993**, *48*, 14936−14943.

(52) Mayadas, A. F.; Shatzkes, M. Electrical-Resistivity Model for Polycrystalline Films: The Case of Arbitrary Reflection at External Surfaces. *Phys. Rev. B* **1970**, *1*, 1382−1389.

(53) Chambers, R. G.; Frisch, O. R. The Anomalous Skin Effect. *Proc. R. Soc. Lond. A* **1952**, *215*, 481−497.

## ■ NOTE ADDED AFTER ASAP PUBLICATION

Due to a production error, this paper was published on the Web on June 18, 2020, with errors in Table 1, column 1, row 3. The corrected version was reposted on June 19, 2020.